\documentclass[prd,
 reprint,
 amsmath,amssymb,
 aps,times
]{revtex4-2}

\usepackage{amsmath}
\usepackage{booktabs} 
\usepackage{threeparttablex}
\usepackage{graphicx}% Include figure files
\usepackage{dcolumn}% Align table columns on decimal point
\usepackage{bm}% bold math
\usepackage{amssymb}
\usepackage{times}
\usepackage{threeparttable}

\def\apj{The Astrophysical Journal}
\def\apjl{The Astrophysical Journal Letter}
\def\apjs{The Astrophysical Journal Supplement}

\def\aap{Astronomy and Astrophysics}

\def\nat{Nature}

\usepackage{orcidlink}
\begin{document}

\preprint{APS/123-QED}

\title{Can Quasi-periodic Eruptions Produce Detectable High Energy Neutrinos?}% Force line breaks with \\
% \thanks{A footnote to the article title}%

\author{Zhi-Peng Ma\,\orcidlink{0009-0007-0717-3667}}
\affiliation{Department of Astronomy, School of Physics, Huazhong University of Science and Technology, Wuhan, Hubei 430074, China}

\author{Kai Wang\,\orcidlink{0000-0003-4976-4098}}
\email{kaiwang@hust.edu.cn}
\affiliation{Department of Astronomy, School of Physics, Huazhong University of Science and Technology, Wuhan, Hubei 430074, China}%

\date{\today}% It is always \today, today,
             %  but any date may be explicitly specified

\begin{abstract}
Quasi-periodic eruptions (QPEs) are a class of X-ray flaring phenomena that occur at the centers of galactic nuclei and are likely to arise from repeated interactions between a star and an accretion disk. This work investigates whether such disk-crossing events can accelerate protons and generate detectable high-energy neutrinos. Based on observed QPE luminosities, recurrence periods, and flare durations, the stellar motion parameters and the disk properties are evaluated. We consider proton acceleration during the breakout phase and evaluate neutrino production, accounting for both $pp$ and $p\gamma$ interactions. Applying the method to ten observed QPE sources, we estimate the neutrino fluence accumulated over a 10-year observation period and compute the corresponding detection numbers for IceCube and IceCube-Gen2. Our analysis indicates that protons can be accelerated up to several tens of TeV, and neutrino production is mostly confined below $\sim 10~\mathrm{TeV}$. The resulting optimized neutrino fluence spans from \(7.0 \times 10^{-7}\) to \(1.5 \times 10^{-4}~\mathrm{GeV~cm^{-2}}\) for these ten QPE sources. We find that the expected neutrino detection number for a single QPE source is low, and the expected neutrino detection number would approach unity only for the most promising QPE source occurring at a distance closer than a few Mpc. Next-generation neutrino telescopes with better detection sensitivities at $\lesssim \rm TeV$ can significantly improve the capture capacity of the cumulative neutrino signal from the QPE population.
\end{abstract}

\maketitle

%\tableofcontents

\section{Introduction}
\label{sec:intro}

Quasi-Periodic Eruptions (QPEs) are recurring X-ray bursts observed at the centers of low-mass galactic nuclei ($10^{5-6.5}M_{\odot}$). QPE flare can last $0.5-10$ hours, reach luminosities of $10^{41-44} \, \text{erg/s}$, exhibiting thermal-like spectra with a temperature of $\sim$100 eV~\citep{miniutti2023alive,giustini2020x,chakraborty2021possible,chakraborty2025discovery,quintin2023tormund,nicholl2024quasi}. Each flare is followed by a quiescent phase lasting $10-50$ hours, during which the emission is well described by an exposed accretion disk~\citep{van2019late,mummery2024fundamental}. The origin of flares is commonly attributed to accretion disk instabilities~\citep{pan2022disk,pan2023application,kaur2023magnetically,sniegowska2023modified,raj2021disk}, mass transfer from the orbiting body onto the central black hole (BH)~\citep{king2020gsn,zhao2022quasi,krolik2022quasiperiodic,zalamea2010white,lu2023quasi,wang2022model,metzger2022interacting}, or the interaction between the disk and an orbiting stellar-mass object~\citep{dai2010quasi,xian2021x,sukova2021stellar,franchini2023quasi,linial2023emri+,linial2024ultraviolet,vurm2025radiation}. The last scenario has been extensively explored, wherein a stellar-mass object in an extreme mass-ratio inspiral (EMRI) periodically intersects the accretion disk formed by tidal disruption event (TDE), generating shocks and ejecting material during each passage and producing the observed flares. For the TDE disk+EMRI scenario, two pieces of observational evidence have been identified. One is that QPEs have been detected from some TDE sources with a $\sim 1$ year time delay~\citep{2024Natur.634..804N,bykov2024further,2025ApJ...983L..39C,2025ApJ...983L..18J}. Another is the alternating long-short pattern in the recurrent QPE time with a long- and short-recurrent time, that is, $T_{\rm long}(t)$ and $T_{\rm short}(t)$, while the sum of long- and short-recurrent time, that is, $T_{\rm long}(t)+T_{\rm short}(t)$ remains constant during the QPE~\citep{PhysRevD.109.103031,2025arXiv250411078Z}.

\begin{figure*}[t]
\centering
\includegraphics[width=\textwidth]{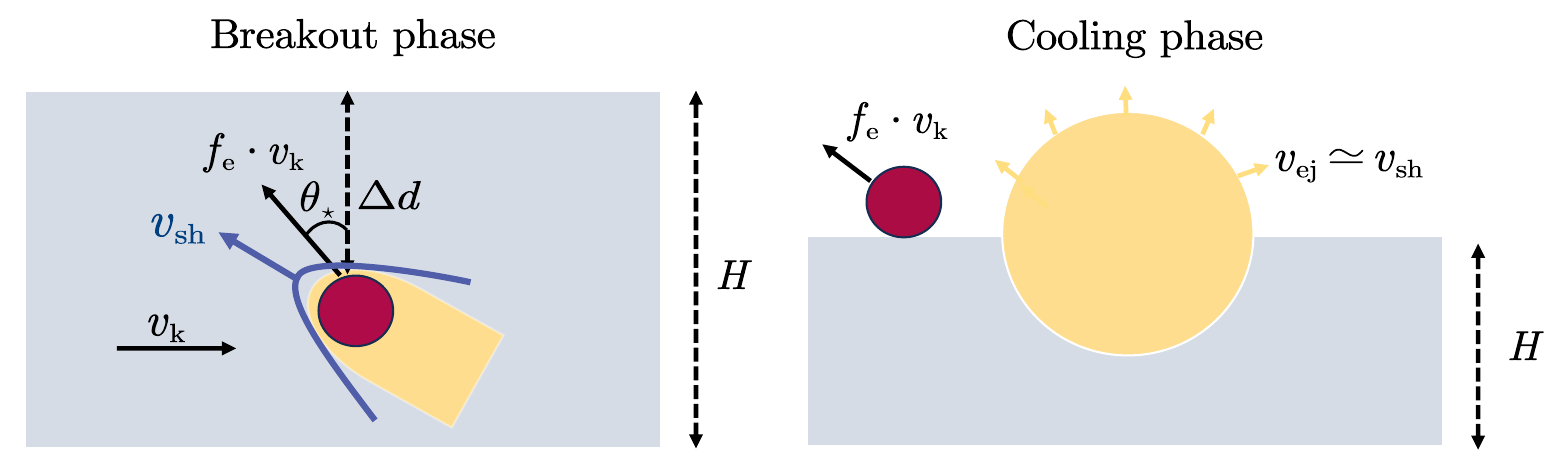}
\caption{A schematic illustration (not to scale) of the proposed model. \textit{Left}: As the star traverses the disk, a radiation-mediated bow shock forms ahead of it. Photons trapped in the shocked gas are released near the disk edge at the breakout height, producing breakout emission. After the shock breakout, particle acceleration and neutrino production start to occur. \textit{Right}: Following the breakout, the shocked gas expands as ejecta, giving rise to the cooling emission.}
\label{fig:sketch}
\end{figure*}

While QPEs have primarily been studied as electromagnetic transients, their dynamic and extreme environments may also enable non-thermal particle acceleration, potentially leading to neutrino production. In particular, QPEs are likely associated with TDEs, and TDEs have long been proposed as plausible sources of high-energy neutrinos~\citep {wang2016tidal,zheng2023choked,yang2024choked,wu2022could,murase2020high}. For QPEs, in the framework of the star-disk collision model (i.e., the TDE disk+EMRI scenario), the shock formed during the star–disk interaction may transition to a collisionless regime near the disk edge, allowing for efficient proton acceleration and subsequent high-energy neutrino production through hadronuclear ($pp$) and photohadronic ($p\gamma$) processes~\citep{waxman2001tev,kaaz2023hydrodynamic,murase2011new}. Such a process is similar to the interactions of stars crossing the AGN disks~\citep{tagawa2023flares,2024ApJ...960...69W,Fan_2024}. The detection of neutrinos from QPEs would provide crucial insights into their underlying physical mechanisms, offering a valuable probe of the high-energy processes inside.

In this work, we investigate the potential of QPEs as sources of high-energy astrophysical neutrinos. Utilizing the information encoded in observed electromagnetic signals, we develop an analytical method to constrain stellar motion parameters and infer disk properties based solely on observables such as luminosity, duration, and recurrence period, without assuming a specific disk model. These constraints then enable us to evaluate the resulting neutrino production.

This paper is organized as follows. In Sec.~\ref{sec:model}, we present the derivation and formulation of our analytical framework. The method for estimating neutrino production is detailed in Sec.~\ref{sec:neu}. In Sec.~\ref{sec:source}, we apply our model to ten observed QPE sources. A summary and discussion of the results are provided in Sec.~\ref{sec:sum}.

\section{Star \& Disk Interaction} \label{sec:model}
We begin by describing the interaction between the star and the disk, as illustrated in Fig.~\ref{fig:sketch}. The star traverses the disk along a nearly linear trajectory with a velocity \( v_{\star} = f_{\rm e} \cdot v_{\rm K} \), where \( v_{\rm K} \) is the Keplerian velocity and \( f_{\rm e}\in [0,2) \) is a factor dependent on the orbital eccentricity \( e \), where $\min f_{\rm e}=(1-e)$ and $\max f_{\rm e}=(1+e)$. Previous studies suggest that the star intersects the disk twice per orbital period \( P_{\rm orb} \), generating one QPE flare at each crossing~\citep{linial2023emri+}. Thus, the QPE recurrence period is \( P_{\rm QPE} = P_{\rm orb}/2 \). Given the mass of the central BH, the stellar velocity \( v_{\star} \) can then be expressed as
\begin{equation}\label{eq:vel}
    v_{\star} 
    = f_{e} \left( \frac{G M_{\rm BH} \pi}{P_{\rm QPE}} \right)^{1/3}
    \simeq 0.08\,c \cdot f_{e} M_{{\rm BH},6}^{1/3} 
    \left( \frac{P_{\rm QPE}}{10{\rm h}} \right)^{-1/3}
\end{equation}

where $M_{\rm {BH},6}=M_{\rm {BH}}/10^6M_{\rm \odot}$, $M_{\rm \odot}$ is the solar mass, and we adopt the conventional notation \( Q_{x} = Q / 10^x \) in cgs units hereafter.

As the star traverses the disk, a radiation-mediated bow shock forms ahead of it~\citep{tagawa2023flares,nayakshin2004x,linial2023emri+}, which heats and compresses the disk gas. Considering the relative motion between the star and the disk gas moving at \( v_{\rm K} \), the shock velocity can be written as $
v_{\rm sh}^{2} = \left( v_{\rm K} + v_{\star} \sin\theta_{\star} \right)^2 + v_{\star}^{2} \cos^2\theta_{\star}
$, where \( \theta_{\star} \) is the angle between the stellar velocity vector and the disk’s angular momentum axis. When the star reaches the breakout height $H_{\rm{bo}}$, photons trapped in the shocked gas begin to escape, producing breakout emission. Simultaneously, the post-shock gas expands as ejecta, giving rise to the cooling emission. The observed QPE emission is expected to comprise both breakout and cooling components, which encode key physical information relevant to neutrino production. In the following, we derive the constraints for necessary physical parameters in terms of observable quantities.

\subsection{Cooling Phase}\label{sec:cooling}
We assume that most of the observed radiation energy is released during the cooling phase (e.g. Ref.~\cite{linial2023emri+,franchini2023quasi}). The spherically expanding ejecta will effectively radiate until the optical depth satisfies~(e.g. Ref.~\cite{levinson2020physics,chen2024electromagnetic})
\begin{equation}\label{eq:tau}
    \tau _{\rm {ej}}(t)\simeq \frac{\kappa M_{\text{ej}}}{4\pi R_{\rm{\rm ej}}^{2}}=\frac{c}{v_{\rm ej}},
\end{equation}
where $\kappa \simeq 0.34~{\rm cm^2~g^{-1}}$ is the opacity due to Thomson scattering, $M_{\rm \rm{ej}}$ is the ejecta mass. The total energy of the ejecta can be estimated as $E_{\mathrm{ej}} \approx M_{\mathrm{ej}} v_{\mathrm{sh}}^2$, roughly half of which goes into kinetic form~\citep{svirski2012optical}, giving $E_{\mathrm{kin}} \approx \frac{1}{2} M_{\mathrm{ej}} v_{\mathrm{ej}}^2 \approx \frac{1}{2} E_{\mathrm{ej}}$, thus $v_{\mathrm{ej}} \sim v_{\mathrm{sh}}$. The expansion of ejecta can be treated approximately as a ballistic expansion. The reason is that, after the breakout site, the thermal-pressure-driven expansion is very short since the shocked expanding layer is highly opaque, as in the supernova case~\citep{chevalier1982self}, which occurs over just a few expansion timescales~\citep{linial2023emri+}. Therefore, following a short acceleration phase, the ejecta enters ballistic expansion with a nearly constant velocity $v_{\rm ej}$.

The ejecta radius $R_{\rm {ej}}$ can be estimated as $R_{\rm ej}\simeq v_{\rm {ej}}\cdot t_{\rm{QPE}}$, where $t_{\rm {QPE}}$ is the duration time of one QPE flare~\citep{tagawa2023shock}. From Eq (\ref{eq:tau}), we can express $M_{\rm {ej}}$ as
\begin{equation}\label{eq:mass_1}
M_{\text{ej}}=\frac{4\pi c}{\kappa}v_{\rm{sh}}t_{\rm{QPE}}^{2}\simeq 5\times 10^{-6}M_{\odot}\,v_{\text{sh,-1}}\left( \frac{t_{\rm QPE}}{0.5{\rm h}} \right)^{2},
\end{equation}
where $v_{\rm{sh,-1}}=v_{\rm{sh}}/0.1c$. The total energy contained in ejecta is
\begin{equation}
\label{eq:ej_enery}
\begin{aligned}
    E_{\rm ej} 
    &\simeq M_{\rm ej} v_{\rm sh}^{2} 
     = \frac{4\pi c}{\kappa} v_{\rm sh}^{3} t_{\rm QPE}^{2} \\
    &\simeq 9.7 \times 10^{46} \ v_{\rm sh,-1}^{3} 
\left( \frac{t_{\rm QPE}}{0.5{\rm h}} \right)^{2}~{\rm erg}.
\end{aligned}
\end{equation}
%This energy expression is analogous to that of a supernova~\citep{svirski2012optical}.
This value is comparable to the result obtained from the analysis of orbital energy loss~\citep{zhou2025secular}.
As mentioned above, half of $E_{\rm ej}$ is converted into kinetic energy to drive the ejecta expanding (due to radiation pressure), while the other half remains as thermal energy in the form of radiation. Since the thermal energy undergoes adiabatic losses before being radiated, we have a relation between observed QPE energy $E_{\rm QPE}$ and $E_{\rm ej}$, i.e.
\begin{equation}\label{eq:energy}
    E_{\rm QPE}=\frac{1}{2}E_{\rm ej}\left( \frac{V_0}{4\pi v_{\rm ej}^{3}t_{\rm QPE}^{3}} \right) ^{1/3},
\end{equation}
where \( V_0 \simeq \pi R_{\star}^{2} H_{\rm bo} / (7 \cos\theta_{\star}) \) is the initial volume of the shocked gas at breakout location, \( \pi R_{\star}^2 \) represents the stellar cross-section, and the factor of 7 accounts for the compression ratio of a radiation-mediated shock (e.g.~Ref.~\cite{kaaz2023hydrodynamic,mabey2020calculating}). The observed QPE energy is estimated as \( E_{\rm QPE} \approx L_{\rm QPE} \cdot t_{\rm QPE} / 2 \), where \( L_{\rm QPE} \) is the peak X-ray luminosity of a single flare. Given the scarcity of QPE detections in other bands, this approximation of energy is reasonable. From Eq.~(\ref{eq:ej_enery}) and Eq.~(\ref{eq:energy}), the radiation efficiency can be defined as  
\begin{equation}
\eta_{\rm QPE} \simeq 2\%~ v^{-3}_{\rm sh,-1} \left( \frac{t_{\rm QPE}}{0.5~\mathrm{h}} \right)^{-1} L_{\rm QPE,42},
\end{equation}
which is smaller than in Ref.\cite{zhou2025secular} but in the same order of magnitude, where $L_{\rm QPE}=L_{\rm QPE}/10^{42}~{\rm erg~s^{-1}}$.

From Eq (\ref{eq:energy}),  we can write the breakout height as
\begin{equation}\label{eq:height}
\begin{aligned}
    H_{\rm bo} 
    &= \cos\theta_\star \cdot \frac{7\kappa^3}{16\pi^3 c^3} v_{\rm sh}^{-6} L_{\rm QPE}^{3} R_\star^{-2} \\
    &\simeq 3 \times 10^{12}~\cos\theta_\star \cdot v_{\rm sh,-1}^{-6} 
    L_{\rm QPE,42}^{3} R_{\star,11}^{-2}~{\rm cm},
\end{aligned}
\end{equation}
where $R_{\rm \star,11}=R_{\star}/10^{11}{\rm cm}$. Since shock breakout occurs near the surface of the disk, the height of the disk scale can be approximated as $ H \approx H_{\rm bo} $.

We note that that the ejecta mass can also be estimated as $M_{\rm ej} \simeq 2 \rho_{\rm d} \pi R_{\star}^2 H/\cos\theta_{\star}$, representing the mass of the disk gas swept up by the star, where \( \rho_{\rm d} \) is the disk density. Equalizing this mas equation with Eq~(\ref{eq:mass_1}), we can write $\rho_{\rm d}$ as
\begin{equation}\label{eq:rho}
    \begin{aligned}
	\rho _{\rm d}&=\frac{32\pi ^3c^4}{7\kappa ^4}v_{\rm sh}^{7}L_{\rm QPE}^{-3}t_{\rm QPE}^{2}\\
	&\simeq 6\times 10^{-8}\,v_{\rm sh,-1}^{7}L_{\rm QPE,42}^{-3}\left( \frac{t_{\rm QPE}}{0.5{\rm h}} \right)^{2}{\rm g~cm^{-3}}.
\end{aligned}
\end{equation}
Equations~(\ref{eq:height}) and~(\ref{eq:rho}) imply that, to reproduce the observed flare with peak luminosity \( L_{\rm QPE} \) and duration \( t_{\rm QPE} \), the disk properties ($H$, $\rho_{\rm d}$) and stellar motion ($f_{e}$, $\theta_{\star}, R_{\star}$) must satisfy these two constraints.
\begin{table*}[htbp]
\centering
\begin{threeparttable}
\caption{Observed properties for QPEs, references are listed in notes.}
\label{Tbl：obs}
\begin{tabular}{lccccccc}
\toprule
Source Name & Redshift & RA (degree)& Dec (degree) & $M_{\rm bh}(M_{\odot})$ & $L_{\rm QPE}~({\rm erg~s^{-1}})$ & $t_{\rm QPE}$~(h) & $P_{\rm QPE}$~(h) \\
\midrule
GSN 069\tnote{a}   & 0.0182&19.79 & -34.19& $4.0\times10^{5}$  & $5.0\times10^{42}$  & 1.00 & 8.8 \\
RXJ 1301.9\tnote{b}  & 0.0237 &195.49& 27.78 & $1.8\times10^{6}$  & $1.0\times10^{42}$  & 0.50 & 4.6 \\
eRO-QPE1\tnote{c}  & 0.0500  & 37.95 &-10.33  & $9.1\times10^{5}$  & $1.0\times10^{43}$  & 7.60 & 18.5 \\
eRO-QPE2\tnote{c}    & 0.0175 & 38.70& -44.32 & $2.3\times10^{5}$  & $1.0\times10^{42}$  & 0.45 & 2.4 \\
eRO-QPE3\tnote{d}    & 0.0240 &35.90 & -28.77  & $5.2\times10^{6}$  & $5.0\times10^{41}$  & 2.30 & 18.5 \\
eRO-QPE4\tnote{d}    & 0.0440 &71.39 &-10.20 & $1.7\times10^{7}$  & $6.0\times10^{42}$  & 0.50 & 13.0 \\
XMMSL1\tnote{e}      & 0.0186 &42.32& -4.21 & $8.5\times10^{4}$  & $3.0\times10^{41}$  & 0.30 & 2.5 \\
AT2019qiz\tnote{f}   & 0.0151 &71.66 &-10.23 & $2.5\times10^{6}$  & $1.8\times10^{43}$& 9.00   & 48.4 \\

AT2022upj\tnote{g}   & 0.0540 &5.99 &-14.42 & $1.0\times10^{6}$  & $5.0\times10^{42}$& 14.40   & 48.0\\

Ansky\tnote{h}   & 0.0240 &203.83 &7.47 & $1.0\times10^{6}$  & $2.0\times10^{43}$& 36.00  & 108.0\\
\bottomrule
\end{tabular}
\begin{tablenotes}
\footnotesize
\item[a] \citet{miniutti2019nine}
\item[b] \citet{giustini2020x}
\item[c] \citet{arcodia2021x,chen2022milli}
\item[d] \citet{arcodia2024more}
\item[e] \citet{chakraborty2021possible}
\item[f] \citet{nicholl2024quasi}
\item[g] \citet{chakraborty2025discovery}
\item[h] \citet{hernandez2025discovery}
\end{tablenotes}
\end{threeparttable}
\end{table*}
\subsection{Breakout Phase}
Near the disk edge, the shock dynamical timescale is given by $t_{\rm dyn}\simeq \Delta d/(v_{\star}\cos\theta _{\star})$, where $\Delta d$ is the distance to the disk edge. Shock breakout occurs when $t_{\rm dyn}$ is comparable to the photon diffusion timescale~\citep{kimura2021outflow}, i.e., $t_{\rm dyn}\approx t_{\rm diff}=\Delta d\cdot \tau/c$, where $\tau\simeq\kappa\rho_{\rm d}\Delta d$ is the optical depth to the disk edge. This yields $\Delta d=c/(\kappa\rho _{\rm d}v_{\star}\cos\theta_{\star} )$. Substituting into $t_{\rm dyn}$, we have
\begin{equation}\label{eq:tdyn}
\begin{aligned}
    t_{\rm dyn}&\simeq \frac{c}{\kappa\rho _{\rm d}v^2_{\star}\cos^2\theta _{\star}}\\
    &\simeq 15\,\cos^{-2}\theta _{\star}\rho _{\rm d,-9}^{-1}\left( \frac{v_{\star}}{0.08c} \right)^{-2}~{\rm s},
\end{aligned}
\end{equation}
where $\rho _{\rm d,-9}=\rho_{\rm d}/10^{-9}{\rm g~cm^{-3}}$. This timescale also characterizes the duration of neutrino production and breakout emission, which is significantly shorter than the cooling emission timescale $t_{\rm QPE}$. The shock kinetic luminosity $L_{\rm kin}$ is
\begin{equation}\label{eq:Lkin}
    \begin{aligned}
    L_{\rm kin}&\simeq \pi R_{\star}^{2}\rho _{\rm d}v_{\rm sh}^{3}\\
	&\simeq 5\times 10^{43}\,v_{\rm sh,-1}^{10}L_{\rm QPE,42}^{-3}R_{\star ,11}^{2}\left( \frac{t_{\rm QPE}}{0.5{\rm h}} \right)^{2}~{\rm erg~s^{-1}},
    \end{aligned}
\end{equation}
which corresponds to the breakout luminosity~\citep{tagawa2023shock}, i.e., $L_{\rm bo}=L_{\rm kin}$. The breakout photon energy density is approximated as $u_{\gamma}\simeq \frac{1}{2}E_{\rm ej}/V_0$. Assuming thermal equilibrium, the characteristic photon energy is
\begin{equation}
    \begin{aligned}
	E_{\rm bo}=k_{\rm b}T_{\rm bo}&=k_b\left( \frac{u_{\gamma}}{a} \right) ^{1/4}\\
	&\simeq 407\,v_{\rm sh,-1}^{9/4}L_{\rm QPE,42}^{-3/4}\left( \frac{t_{\rm QPE}}{0.5{\rm h}} \right)^{2}~{\rm eV},
\end{aligned}
\end{equation}
which is also at X-ray band, where $k_{\rm b}$ is Boltzman coefficient and $a$ is blackbody coefficient.

We note that, under the characteristic values adopted above, the breakout luminosity ($L_{\rm bo}=L_{\rm kin}$) exceeds the observed peak luminosity ($L_{\rm QPE,42}$). Furthermore, the energy released during the breakout phase, approximately $7.5 \times 10^{44}~\mathrm{erg}$, is comparable to that released during the cooling phase. This challenges our initial assumption that the radiative energy is dominated by the cooling phase. To maintain consistency with both observation and our assumption, we require that the breakout luminosity does not exceed the observed QPE peak luminosity,
\begin{equation}\label{ineq:L}
    L_{\rm bo} \leq L_{\rm QPE},
\end{equation}
and the breakout timescale remains shorter than the flare duration,
\begin{equation}\label{ineq:t}
    t_{\rm dyn} < t_{\rm QPE}.
\end{equation}
%These conditions impose constraints on the stellar motion, specifically on the parameters \( f_{e} \) and \( \theta_{\star} \).

In summary, the constraint relations, i.e., Eqs.~(\ref{eq:height}), (\ref{eq:rho}), (\ref{ineq:L}), and (\ref{ineq:t}), connect the observed QPE properties \( L_{\rm QPE} \), \( t_{\rm QPE} \), and \( P_{\rm QPE} \) to the model parameters. In our neutrino production calculations, the stellar motion parameters (\( f_{e} \), \( \theta_{\star} \), \( R_{\star} \)) are treated as free variables, constrained by Eqs.~(\ref{ineq:L}) and (\ref{ineq:t}). Given a specific set of these parameters, the corresponding disk properties (\( H \), \( \rho_{\rm d} \)) are then determined via Eqs.~(\ref{eq:height}) and (\ref{eq:rho}).

% By adopting specific observational values, these relations constrain the stellar motion parameters (\( f_{e} \), \( \theta_{\star} \), \( R_{\star} \)) and determine the corresponding disk properties (\( H \), \( \rho_{\rm d} \)).

\section{Neutrino Production} \label{sec:neu}
Before the breakout, the shock is radiation-mediated, the accelerated particles will lose their energies through frequent collisions to the Maxwellian distribution, so that the effective particle acceleration and the non-thermal particle distribution are prohibited \citep{PhysRevLett.87.071101}. After the breakout, photons begin to diffuse out, and the shock becomes collisionless and the effective particle acceleration begins~\citep{murase2014probing,murase2024interacting}. Within the dynamical timescale $t_{\rm dyn}$ following the shock breakout, protons can be efficiently accelerated and subsequently produce high-energy neutrinos. We assume that accelerated protons spectrum follow $dN_{\rm p}/dE_{\rm p} \propto E^{-2}_{\rm p}\exp(-E_{\rm p}/E_{\rm p,max})$, where $E_{\rm p}$ is the proton energy, $E_{\rm p,max}$ is the maximum proton energy~\citep{blandford1987particle,malkov2001nonlinear}. The acceleration timescale is
\begin{equation}
    t_{\rm acc}=\frac{\eta E_{\rm p}}{q_{\rm e}Bc}\simeq0.08~v^{-3}_{\rm sh,0.1}\rho^{-1/2}_{\rm d,-9}\left( \frac{E_{\rm p}}{50~{\rm TeV}} \right)~{\rm s}
\end{equation}
where $\eta\simeq20c^2/3v^2_{\rm sh}$ is for Bohm limit and $q_{\rm e}$ is the electron charge. The magnetic field strength can be estimated by $B=\sqrt{8\pi\epsilon_{\rm B}\rho_{\rm d}v^2_{\rm sh}}$ and $\epsilon_{\rm B}$ is the magnetic field energy fraction. We adopt $\epsilon_B = 0.01$ as a representative value for shock environments, based on observational modeling of young supernova remnants, where typical values range from $10^{-3}$ to $10^{-1}$ \citep{reynolds2021efficiencies}. Similar ranges have also been found in GRB afterglow fits \citep{panaitescu2001fundamental}.

We consider both \( pp \) interactions with the shocked gas and photohadronic \( p\gamma \) interactions with breakout photons. The $pp$ process timescale is
\begin{equation}
t_{\rm pp}=\frac{1}{0.5\sigma_{\rm pp}n_{\rm s}c}\simeq 1\,\rho^{-1}_{\rm d,-9}~{\rm s}, 
\end{equation}
where $\sigma_{\rm pp}\sim5\times10^{-26} {\rm cm}^{-2}$ is the cross section for $pp$ process~\citep{particle2004ata}, $n_{s}=4\rho_{\rm d}/m_{\rm p}$ is the number density of shocked gas~\citep{stecker1968effect,murase2007high}, $m_{\rm p}$ is the proton mass.
The timescale for the photohadronic (\(p\gamma\)) interaction is given by
\begin{equation}
t^{-1}_{p\gamma} = \frac{c}{2\gamma_p^2} \int_{\tilde{E}_{\rm th}}^{\infty} d\tilde{E} \, \sigma_{p\gamma}(\tilde{E}) \kappa_{p\gamma}(\tilde{E}) \tilde{E} \int_{\tilde{E}/2\gamma_p}^{\infty} dE_{\gamma} \, E_{\gamma}^{-2} \frac{dN_{\gamma}}{dE_{\gamma}},
\end{equation}
where $ \sigma_{p\gamma} $ and $ \kappa_{p\gamma} $ are the cross-section and inelasticity, respectively~\citep{stecker1968effect,patrignani2016review}, $ \tilde{E} $ is the photon energy in the proton rest frame, and $ \tilde{E}_{\rm th} \simeq 145~\mathrm{MeV} $ is the threshold energy. Here, $ \gamma_p = E_p / (m_pc^2) $ is the proton Lorentz factor, and $ dN_{\gamma}/dE_{\gamma} $ is the differential number density of the breakout photon. The Bethe–Heitler (BH) process timescale $ t_{\rm BH} $ is calculated using the same expression, with $ \sigma_{p\gamma} $ and $ \kappa_{p\gamma} $ replaced by the corresponding quantities $ \sigma_{\rm BH} $ and $ \kappa_{\rm BH} $~\citep{chodorowski1992reaction}. 
\begin{figure*}[t]
% \begin{center}
\includegraphics[width=0.46\linewidth,height=0.4\linewidth]{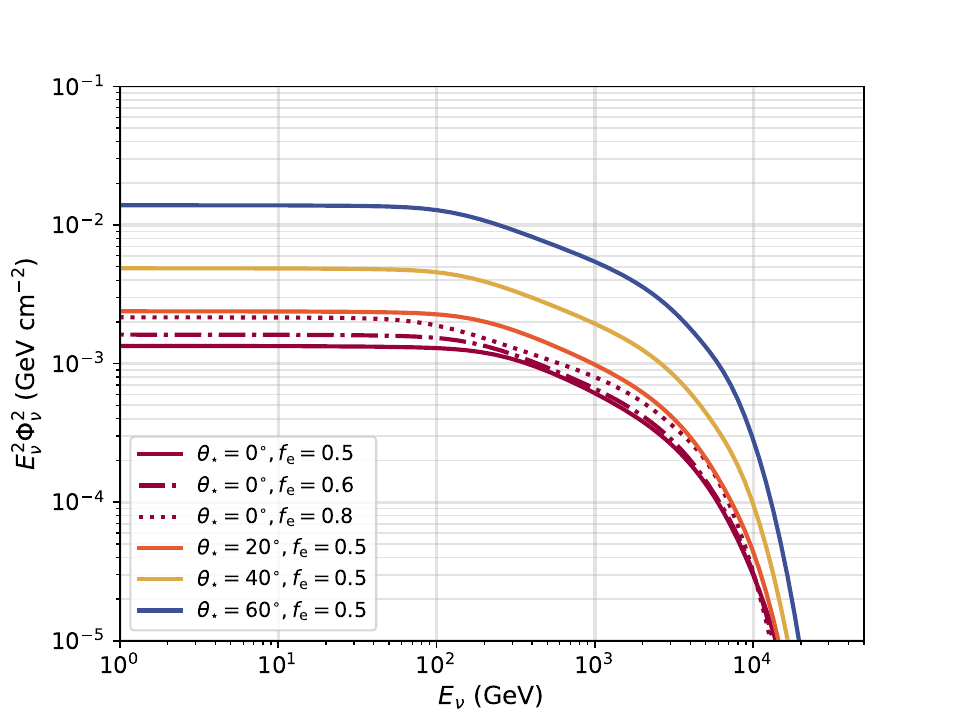}
\includegraphics[width=0.46\linewidth,height=0.4\linewidth]{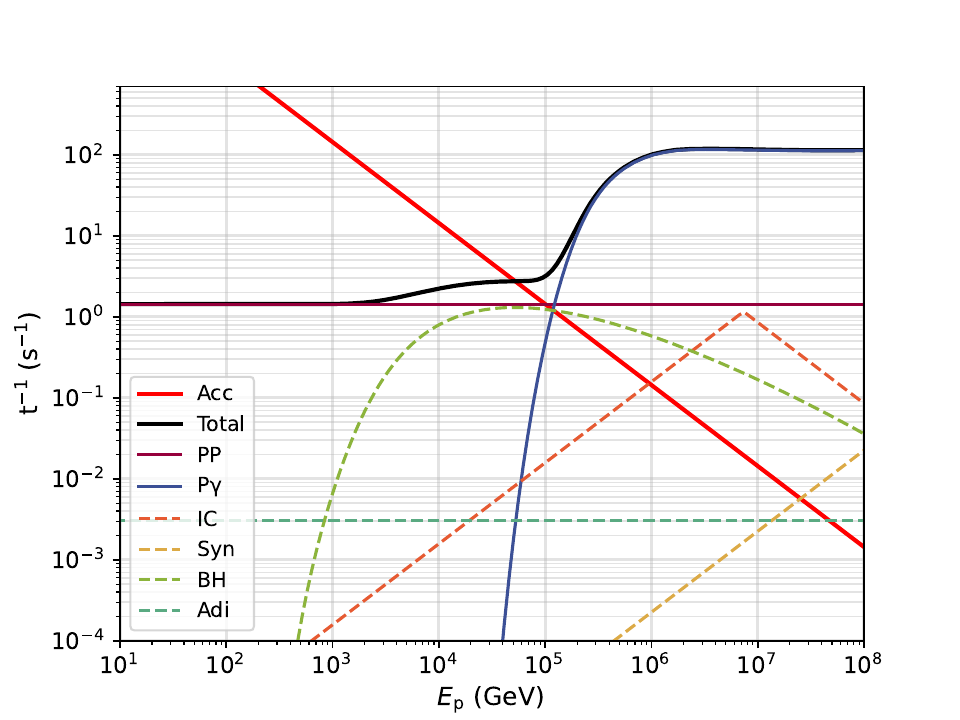}
\caption{ \textit{Left}: The all-flavor neutrino fluence accumulated over 8760 disk crossings, corresponding to a 10-year QPE activity with $P_{\rm QPE} = 10~{\rm h}$. Parameters: $M_{\rm bh} = 10^6 M_{\odot}$, $D_{\rm L} = 100~{\rm Mpc}$, $R_{\star} = 10^{11}~{\rm cm}$, $L_{\rm QPE} = 10^{42}~{\rm erg~s^{-1}}$, and $t_{\rm QPE} = 0.5~{\rm h}$. All adopted ($\theta_{\star}$, $f_{\rm e}$) combinations satisfy Eqs.~(\ref{ineq:L}) and~(\ref{ineq:t}). 
\textit{Right}: Timescale illustration for $\theta_{\star} = 60^{\circ}$ and $f_{\rm e} = 0.5$.}
%\vspace{-1.\baselineskip}
\label{fig:time}
\end{figure*}
Using approximate cross-sections of \( \sigma_{p\gamma} \sim 0.7 \times 10^{-28}~{\rm cm^2} \) and \( \sigma_{\rm BH} \sim 0.8 \times 10^{-30}~{\rm cm^2} \)~\citep{murase2016hidden}, the corresponding timescales under characteristic parameters are \( t_{p\gamma} \simeq [\sigma_{p\gamma}n_{\gamma}c]^{-1} \sim 1.8\times10^{-4}~{\rm s} \) and \( t_{\rm BH} \simeq [\sigma_{\rm BH}n_{\gamma}c]^{-1} \sim 10^{-2}~{\rm s} \), both processes are more efficient than $pp$ interaction, where photon number density $n_{\gamma}\simeq 6\times10^{21}~{\rm cm^{-3}}$, under characteristic parameters. The typical proton energy interacts with the breakout photon for $p\gamma$ process is $E_{p\gamma}\approx0.5m_{\rm p}c^2\tilde{E}_{\Delta}/E_{\rm bo}\simeq 3.5\times10^5~(E_{\rm bo}/400~{\rm eV})^{-1}~{\rm GeV}$, where $\tilde{E}_{\Delta}\approx0.3~{\rm GeV}$. Similarly, for BH process, we have $E_{\rm BH}\approx0.5m_{\rm p}c^2\tilde{E}_{\rm BH}/E_{\rm bo}\simeq 1.2\times10^4~(E_{\rm bo}/400~{\rm eV})^{-1}~{\rm GeV}$, where $\tilde{E}_{\rm BH}\approx10~{\rm MeV}$~\citep{chodorowski1992reaction,stepney1983numerical}. The rough estimation above suggests that the ${p\gamma}$ process dominates over the $pp$ process only above $\sim 3.5\times10^5$ GeV, which is typically beyond the accelerated maximum energy (tens of TeV), the hardronic process of QPEs should be primarily governed by $pp$, while the BH process may suppress neutrino production in the lower energy band.

Protons also undergo energy losses via adiabatic expansion, synchrotron radiation, and inverse Compton (IC) scattering. The adiabatic cooling timescale is approximately the dynamical timescale, \( t_{\rm adi} \simeq t_{\rm dyn} \). Synchrotron radiation cooling has a timescale
\begin{equation}
\begin{aligned}
        t_{\rm syn}&=\frac{6\pi m^4_{\rm p}c^3}{\sigma_{\rm T}m^2_{\rm e}B^2E_{\rm p}}\\
        &\simeq4\times 10^4 ~v^{-1}_{\rm sh,0.1}\rho^{-1/2}_{\rm d,-9}\left( \frac{E_{\rm p}}{50~{\rm TeV}} \right)^{-1}~{\rm s},
\end{aligned}
\end{equation}
which is inefficient, where $\sigma_{\rm T}$ is Thomson cross-section, and $m_{\rm e}$ is the electron mass.
The timescale of inverse-Compton scattering is~\citep{denton2018exploring}:

\begin{equation}
	t_{\rm IC}=\begin{cases}
		\frac{3m_{p}^{4}c^3}{4\sigma _Tm_{e}^{2}n_{\gamma}E _{\gamma}E _{\rm p}},&E _{\gamma}E _p<m_{p}^{2}c^4,\\
  \\
		\frac{3E _{\gamma}E _p}{4\sigma _Tm_{e}^{2}n_{\gamma}c^5}\,\,  ,&E _{\gamma}E _p>m_{p}^{2}c^4.\\
	\end{cases}
\end{equation}
The IC timescale at break energy $m^2_{\rm p}c^4$ is typically $\sim 0.01~{\rm s}$, which is comparable to that of BH process.

The production of neutrinos can be quantitatively assessed by analyzing the fractional energy loss of protons. The total cooling timescale for protons is defined as $t^{-1}_{\rm cool}=t^{-1}_{\rm pp}+t^{-1}_{\rm p\gamma}+t^{-1}_{\rm BH}+t^{-1}_{\rm syn}+t^{-1}_{\rm IC}+t^{-1}_{\rm adi}$, the total energy loss fraction of protons is $\eta_{\rm p}={\min}(1,t_{\rm dyn}/t_{\rm cool})$.
The fractions of energy loss by $pp$ ($p\gamma$) is estimated by $\xi_{\rm (pp,p\gamma)}=t^{-1}_{\rm (pp,p\gamma)}/t^{-1}_{\rm cool}$. Given the energy loss timescale, the neutrino fluence from a single disk-crossing event can be estimated based on the initial proton luminosity~\citep{wang2009prompt,murase2008prompt,murase2016hidden}
\begin{equation}\label{eq:flc}
    E_{\nu}^{2}\Phi _{\nu ,\left\{ {\rm pp,p\gamma} \right\}}=\frac{1}{4\pi D_{L}^{2}}\frac{3K}{4\left( 1+K \right)}\frac{\eta _p\xi_{\rm (pp,p\gamma)}L_{\rm p}t_{\rm dyn}}{\ln\left( E_{\rm p,max}/1~{\rm GeV} \right)},
\end{equation}
where the neutrino energy $E_{\nu}\simeq0.05E_{\rm p}$, $D_{\rm L}$ is the luminosity distance from the QPE source, $K=2~(K=1)$ is for $pp$~($p\gamma$) process, $L_{\rm p}=0.1\cdot L_{\rm kin}$ is the accelerated proton luminosity. The maximum proton energy $E_{\rm p,max}$ is determined by the intersection of $t_{\rm cool}$ and $t_{\rm acc}$.

Equation~(\ref{eq:flc}) applies to a single disk–crossing event; however, the \emph{total} neutrino fluence must be integrated over the QPE’s active lifetime, \(\tau_{\rm QPE}\). Since the lifetime of the QPE activity is highly uncertain, which ranges from a few years (e.g.\ eRO‑QPE2) to decades (e.g.\ RX\,J1301.9+2747)~\citep{arcodia2021x,giustini2020x}, we here adopt a characteristic value of $\tau_{\rm QPE} = 10\ \mathrm{yr}$~\citep{arcodia2024cosmic}.

\begin{table*}[htbp]
\centering
\caption{Optimized neutrino detection numbers ($\nu_{\mu}+\bar{\nu}_{\mu}$) and corresponding free parameters ($\theta_{\star}$, $f_{\rm e}$, $R_{\star}$). The derived disk properties ($H$, $\rho_{\rm d}$) are also listed. The last two columns present the expected event numbers for IceCube (IC) and IceCube-Gen2 (IC-Gen2) in ten years, assuming that the effective area for IC-Gen2 is seven times that of IC. The expected number for IC accounts for contributions from both the IC-86 configuration ($E_{\nu_{\mu}} > 100$ GeV)  and the DeepCore subarray  ($E_{\nu_{\mu}} < 100$ GeV). For comparison, the effective area of DeepCore subarray is also assumed to be seven times larger in the next generation.}
\label{Tbl：neu}
\begin{tabular}{lccccccc}
\toprule
Source Name & $\theta_{\star}$(degree) & $f_{\rm e}$& $R_{\star}$ (cm) & $H$ (cm) & $\rho_{\rm d}~({\rm g~cm^{-3}})$ &$N_{\nu_{\mu}+\bar{\nu}_{\mu}}$~(IC)&$N_{\nu_{\mu}+\bar{\nu}_{\mu}}$~(IC-Gen2)    \\
\midrule
GSN 069   & 80 & 0.8 & $2\times10^{11}$&$2.4\times10^{13}$ &$1.1\times10^{-9}$ & $1.3\times10^{-6}$ & $9.1\times10^{-6}$ \\

RXJ 1301.9   & 75 & 0.3 & $2\times10^{11}$&$1.4\times10^{12}$  &$5.5\times10^{-9}$& $8.5\times10^{-5}$ & $5.9\times10^{-4}$ \\

eRO-QPE1   & 80 & 0.6 & $5\times10^{11}$&$1.5\times10^{14}$ &$1.3\times10^{-9}$& $1.6\times10^{-6}$ & $1.2\times10^{-5}$ \\

eRO-QPE2   & 80 & 0.5 & $1\times10^{11}$&$3.0\times10^{12}$  &$6.1\times10^{-9}$& $1.2\times 10^{-6}$ &$ 8.2\times10^{-6} $ \\
eRO-QPE3   & 70 & 0.15 & $2\times10^{11}$&$3.0\times10^{13}$ &$3.2\times10^{-9}$&$6.3\times10^{-8}$&$4.4\times10^{-7}$\\
eRO-QPE4   & 75 & 0.3 & $4\times10^{11}$&$6.7\times10^{12}$ &$4.4\times10^{-10}$&$1.6\times10^{-6}$&$1.1\times10^{-5}$\\
XMMSL1   & 70& 0.4 & $2\times10^{11}$&$1.3\times10^{12}$ &$1.7\times10^{-9}$&$4.6\times10^{-5}$ &$3.3\times10^{-4}$\\
AT2019qiz   & 75 & 0.6 & $6\times10^{11}$&$8.0\times10^{14}$ &$3.5  \times10^{-10}$&$4.9\times10^{-6}$ &$3.5\times10^{-5}$\\
AT2022upj   & 70 & 0.6 & $2\times10^{11}$&$7.2\times10^{14}$ &$1.0  \times10^{-8}$&$1.6\times10^{-8}$ &$1.1\times10^{-7}$\\

Ansky   & 89 & 1.3 & $1\times10^{11}$&$8.0\times10^{14}$ &$1.7  \times10^{-8}$&$1.7\times10^{-4}$ &$1.4\times10^{-3}$\\
\bottomrule
\end{tabular}
\end{table*}

\begin{figure}
    \centering
    \includegraphics[width=1.0\linewidth,height=0.8\linewidth]{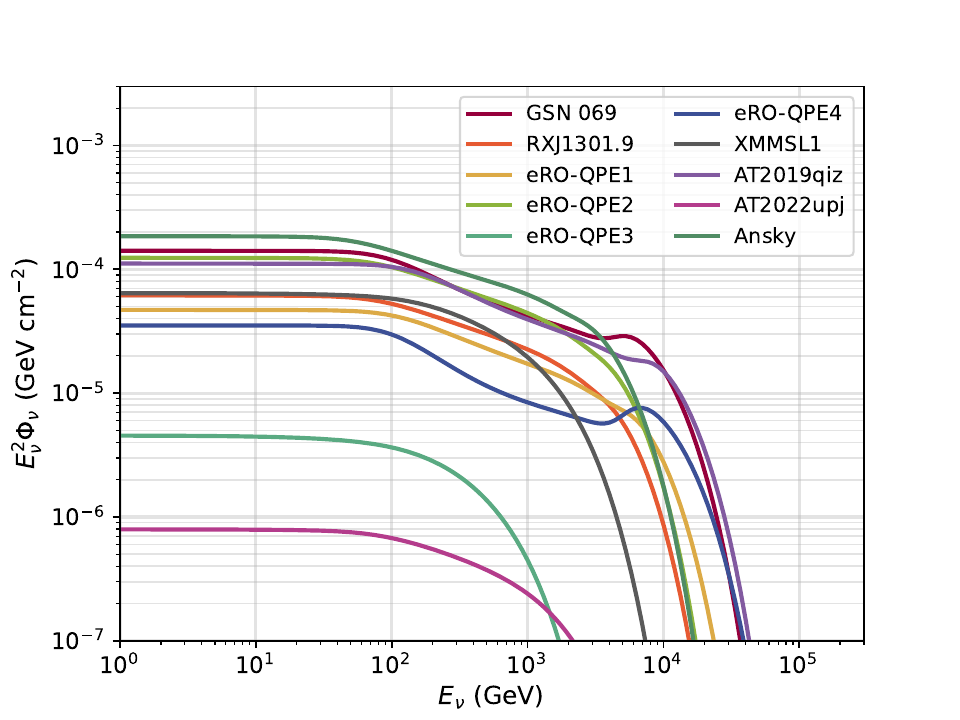}
    \caption{The optimized all-flavor neutrino fluence from ten QPE sources, corresponding to a 10-year QPE activity. The corresponding parameters are listed in Tbl.~\ref{Tbl：neu}.}
    \label{fig:source}
\end{figure}

Fig.~\ref{fig:time} shows the neutrino energy fluence accumulated over 10-year QPE duration—for various combinations of \( f_e \) and \( \theta_{\star} \), assuming $M_{\rm bh}=10^6M_{\odot}$, $D_{\rm L}=100~{\rm Mpc}$, \( R_{\star} = 10^{11}~{\rm cm} \), \( L_{\rm QPE} = 10^{42}~{\rm erg~s^{-1}} \), \( t_{\rm QPE} = 0.5~{\rm h} \), and \( P_{\rm QPE} = 10~{\rm h} \). To satisfy Eqs.~(\ref{ineq:L}) and~(\ref{ineq:t}), it is necessary that $f_{\rm e} < 1$ in all cases, corresponding to a high orbital eccentricity of the star. In general, for fixed \( f_{\rm e} \), a larger \( \theta_{\star} \) yields a higher neutrino fluence, as it implies a later stellar exit from the disk and thus a longer duration of shock interaction with the ambient gas and photon fields. Conversely, for fixed \( \theta_{\star} \), a larger \( f_{\rm e} \) corresponds to a higher shock velocity, also enhancing neutrino production. The resulting neutrino energy fluence under our characteristic parameters ranges from \( 1 \times 10^{-3} \) to \( 2 \times 10^{-2}~{\rm GeV~cm^{-2}} \). We find that \( E_{\rm p, max} \) reaches several tens of TeV for all cases. A representative plot of the relevant timescales is also included in Fig.~\ref{fig:time}.

\section{Neutrinos from QPE Sources}\label{sec:source}
In this section, we apply our analysis method to ten specific QPE sources, whose observed properties are summarized in Tbl.~\ref{Tbl：obs}. To account for both long-lived and recurrent QPE sources, a uniform integration time of $\tau_{\rm QPE} = 10$~yr is adopted in estimating the accumulated neutrino fluence. For each source, we ensure that the adopted ($\theta_{\star}$, $f_{\rm e}$, $R_{\star}$) values satisfy Eqs.~(\ref{ineq:L}) and~(\ref{ineq:t}), and determine the disk properties via Eqs.~(\ref{eq:height}) and~(\ref{eq:rho}). We then search for parameter combinations that maximize the resulting neutrino fluence. Technically, we begin with the maximum values of $\theta_{\star} = 89^\circ$, $f_{\rm e} = 2$, and $R_{\star} = 6 \times 10^{11}~\mathrm{cm}$, and then systematically decrease these parameters, one at a time, in order to identify the combination that maximizes neutrino emission while remaining consistent with observational constraints. Similar to the results in the previous section, for most of the sources, $f_{\rm e} <1 $ is required to satisfy the constraints, and larger values of $\theta_{\star}$ are preferred to enhance neutrino production. 

\begin{figure}
    \centering
\includegraphics[width=1.0\linewidth,height=0.8\linewidth]{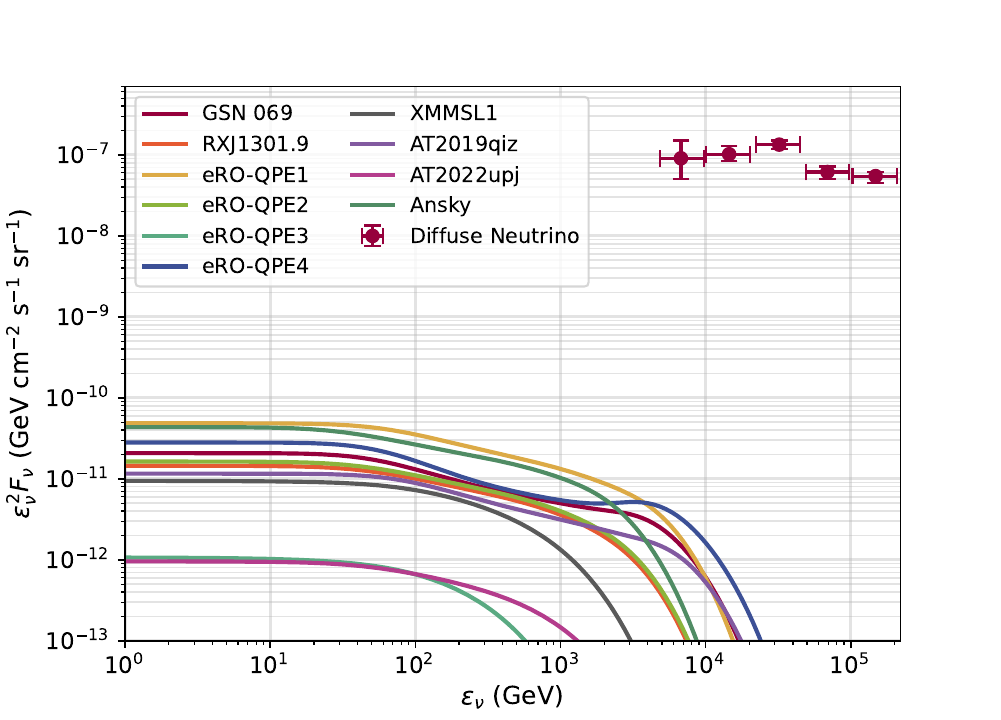}
\caption{Estimated contribution to the diffuse neutrino background. All QPE sources are assumed to be the same as one specific source, which has been shown in the legend. In addition, each source is assumed to have 10 years of QPE activity. The diffuse neutrino background data are taken from Ref.~\cite{naab2023measurement}.}
    \label{fig:diff}
\end{figure}
The results are presented in Fig.~\ref{fig:source}. The predicted neutrino fluence spans from $7.0 \times10^{-7}$ to $1.5 \times 10^{-4}~{\rm GeV~cm^{-2}}$, which remains relatively low, particularly compared to other potential disk-associated neutrino sources such as supernovae or binary black hole mergers within active galactic nuclei (AGNs) disks~\citep{zhu2021thermonuclear,ma2024high,zhou2023high,zhou2023high1}. We find that \( E_{\rm p, max} \) ranges from several TeV to tens of TeV for all sources, and the resulting neutrino production is primarily concentrated below $\sim 10~\mathrm{TeV}$ range and declines sharply at the highest energy. The slight dip between $\sim 100~\mathrm{GeV}$ and $\sim 10~\mathrm{TeV}$ is attributed to the BH process, consistent with the analytical estimates discussed earlier.

We estimate the expected neutrino detection counts ($\nu_{\mu}+\bar{\nu}_{\mu}$) as follows
\begin{equation}
    N_{\nu _{\mu}}=\int_{1 {\rm GeV}}^{100 {\rm PeV}} A_{\rm eff}(E_{\nu _{\mu}})\Phi_{\nu _{\mu}} dE_{\nu _{\mu}},
\end{equation}
where the (anti-)muon neutrino fluence is one-third of the all-flavor neutrino fluence considering the neutrino oscillation, i.e., $\Phi _{\nu _{\mu}}=\frac{1}{3}\Phi _{\nu}$. The effective area, \( A_{\rm eff}(E_{\nu_\mu}) \), includes contributions from both the IC-86 configuration (\( E_{\nu_\mu} > 100~\mathrm{GeV} \),~Ref.~\cite{aartsen2020time}) and the DeepCore subarray (\( E_{\nu_\mu} < 100~\mathrm{GeV} \),~Ref.~\cite{aartsen2017search}), with the IC-86 array further accounting for declination-dependent variations in the effective area~\citep{aartsen2020time}. The results are summarized in Tbl.~\ref{Tbl：neu}, along with the corresponding parameters ($\theta_{\star}$, $f_{\rm e}$, $R_{\star}$, $H$, $\rho_{\rm d}$). 

The predicted number of neutrinos from each QPE source over a 10-year period is on the order of $10^{-7}$ to $10^{-4}$, which is notably low. This is primarily since the neutrino emission from these sources mainly concentrates in $E_\nu < 100~\mathrm{GeV}$, within the DeepCore energy band, where the effective area is much smaller compared to the main IceCube detector. These estimated event rates suggest that the detection of neutrinos from individual QPE sources remains highly challenging.

For the most promising QPE source, i.e., QPE Ansky, the expected neutrino detection numbers are $1.7 \times 10^{-4}$ by IceCube and $1.4 \times 10^{-3}$ by IceCube-Gen2. To guarantee the detection of one neutrino event, for an Ansky-like QPE event, it has to occur at a much closer distance, says, $\sim 1.4 \,\rm Mpc$ for IceCube and $\sim 4 \,\rm Mpc$ for IceCube-Gen2.

Based on the neutrino fluence for an individual QPE, we estimate the QPE contribution to the diffuse neutrino flux via
\begin{equation}
    \epsilon^2_{\nu}F_{\nu} = \frac{c}{H_0} \int_0^{z_{\rm max}} 
    \frac{R_0 f(z) E_\nu^2 \Phi_\nu(E_\nu) D^2_{\rm L}}{(1+z)^2 \sqrt{\Omega_{\rm m}(1+z)^3 + \Omega_\Lambda}} \, dz,
\end{equation}
where $\epsilon_\nu = E_\nu/(1+z)$ is the observed neutrino energy, and $R_0/\tau_{\rm QPE} \approx 0.6 \times 10^{-7}~{\rm Mpc}^{-3}~{\rm yr}^{-1} (\tau_{\rm QPE}/10~{\rm yr})^{-1}$ is the QPE formation rate~\citep{arcodia2024cosmic}. The redshift evolution $f(z)$ is assumed to follow that of TDEs, given by
$f(z) = \left[(1+z)^{0.2\eta} + \left(\frac{1+z}{1.43}\right)^{-3.2\eta} + \left(\frac{1+z}{2.66}\right)^{-7\eta} \right]^{1/\eta}$, where $ \eta = -2$~\citep{sun2015extragalactic}. The resulting contribution to the diffuse neutrino background is found to be $\sim 10^{-12}$–$ 10^{-10}~{\rm GeV}~{\rm cm}^{-2}~{\rm s}^{-1}~{\rm sr}^{-1}$, indicating a negligible contribution to the TeV-PeV diffuse neutrino background, as shown in Fig.~\ref{fig:diff}.

Finally, if QPEs originate from a stellar-mass black hole instead of a star, the effective interaction radius $R_*$ can be approximated by the Bondi radius, i.e., $R_* \approx 3 \times 10^8~\mathrm{cm} \left( \frac{M_{\rm bh}}{10\,M_\odot} \right)v^{-2}_{\rm sh,-1},$ where $M_{\rm bh}$ is the black hole mass. This radius is significantly smaller than the typical stellar radius $\sim 10^{11}$~cm. According to $L_{\rm kin}\simeq \pi R_{\star}^{2}\rho _{\rm d}v_{\rm sh}^{3}$, this would significantly reduce the shock kinetic luminosity and, consequently, the neutrino emission, assuming that other parameters remain unchanged. Note that in Eq.~(\ref{eq:Lkin}) a typical radius of a star without depending on the shock velocity (or approximately the relative velocity of the stellar black hole to the disk gas) is adopted as in Ref.~\citep{linial2023emri+}. However, if an influence radius of the EMRI companion is adopted as suggested in Ref.~\citep{franchini2023quasi}, such as a Bondi radius of a stellar-mass black hole, which presents $R_*\propto v^{-2}_{\rm sh}$, the shock kinetic luminosity will be $L_{\rm kin}\propto v_{\rm sh}^{-1}$. In the case that the motion direction of the stellar-mass black hole is with a small angle with the motion direction of the disk gas (i.e., a small velocity of relative movement or a small shock velocity $v_{\rm sh}$), the Bondi radius can be boosted to be large, and the resulting shock kinetic luminosity will also increase, leading to a large enough energy budget for particle accleration and consequently a strong enough neutrino production.

\section{Summary}\label{sec:sum}

In this work, we explore the potential of QPEs as sources of high-energy neutrinos. The star motion parameters and the disk properties can be constrained and inferred based on observed QPE characteristics such as luminosity, duration, and recurrence period.

We evaluate the neutrino fluence accounting for both $pp$ and $p\gamma$ interactions and apply our method to ten QPE sources. To maximize neutrino production while satisfying the derived constraints, a high orbital eccentricity and a low inclination angle of the stellar orbit are required. Our analysis shows that the maximum proton energy can reach several tens of TeV, and the resulting neutrino production is concentrated primarily below \(10~\mathrm{TeV}\). The corresponding optimized neutrino fluence ranges from \(7.0 \times 10^{-7}\) to \(1.5 \times 10^{-4}~\mathrm{GeV~cm^{-2}}\) for an individual QPE source.

The expected neutrino detection number for IceCube and IceCube-Gen2 over 10 years is estimated, and the resulting detection rate for a single current QPE source is found to be low. The expected neutrino detection for a single QPE source becomes possible only when the QPE occurs within a few Mpc. In the future, if a QPE source occurs at such a distance, it can be used to check the star-disk model of QPE by the neutrino aspect and to study the vicinity of SMBHs.

As mentioned above, QPEs are expected to follow TDEs, and both activities may produce neutrinos. Their origins can be distinguished through the following aspects:
\begin{itemize}
  \item \textbf{Neutrino energy:} Neutrinos from TDEs are expected to be more energetic. All three reported TDE candidates associated with IceCube neutrinos are at $\sim$100 TeV energies~\citep{stein2021tidal,reusch2022candidate,van2024establishing}, while the expected peak energies for QPE-produced neutrinos are below 100 TeV (see Fig.~3).
\item \textbf{Neutrino fluence and duration:} While the fluence of TDE-produced neutrinos is estimated to be $10^{-4}$–$10^{-1}\,\mathrm{GeV\,cm^{-2}}$~\citep{winter2023interpretation}, which is comparable to some QPE cases (see Fig.~3), the TDE emission is likely a short-term flare. In contrast, QPE signals represent cumulative emission over $\sim$10 years.
\item \textbf{Multi-wavelength temporal correlation:} Neutrinos from TDEs are predicted to arrive around $\mathcal{O}(100)$ days after the blackbody peak in optical/UV bands. By this time, the blackbody luminosity has faded, but the infrared dust echo is near maximum~\citep{stein2021tidal,reusch2022candidate,van2024establishing}. QPEs, by contrast, do not show such delayed and multi-band correlated emission.
\end{itemize}

Considering the potential population of QPEs across the sky, the contribution to the TeV-PeV diffuse neutrino background observed by IceCube is also estimated and found to be negligible. Future neutrino detectors with better detection capability at $\lesssim \rm TeV$ will be crucial to capture neutrinos from QPEs.

\begin{acknowledgments}
We thank Prof. Zhen Pan for his very helpful comments and suggestions. We acknowledge support from the National Natural Science Foundation of China under grant No.12003007 and the Fundamental Research Funds for the Central Universities (No. 2020kfyXJJS039).
\end{acknowledgments}

\nocite{*}
\bibliographystyle{apsrev4-2}
\bibliography{reference}

\end{document}